\documentclass[12pt,a4paper]{amsart}
\input xy \xyoption{all} 

\newcommand{\SL}{{\rm SL}}
\newcommand{\sll}{{\rm sl}}   
\newcommand{\SU}{{\rm SU}}

\newcommand{\SO}{{\rm SO}}
\newcommand{\so}{{\rm so}}
\newcommand{\su}{{\rm su}}
\newcommand{\Uq}{{\rm U}_q}
 \newcommand{\dd}{{\rm d}} 

\newcommand{\R}{{\mathbb R}}
\newcommand{\C}{{\mathbb C}}

\newcommand{\nfactor}{{\frac1{2\pi^2}}}
\newtheorem{defn}{Definition}
\newtheorem{conj}{Conjecture} 

\newcommand{\FourX}[4]  
{
\xymatrix{\ar@{-}[dr]^{#1} & & \ar@{-}[dl]^{#2} \\
 & *{\bullet} & \\
\ar@{-}[ur]^{#3} & & \ar@{-}[ul]^{#4}\\}
}

\newcommand{\DoubleY}[5] 
{
\xymatrix{\ar@{-}[dr]^{#1} & & & \ar@{-}[dl]_{#2} \\
 & *{\bullet} \ar@{-}[r]^{#5}& *{\bullet} &\\
\ar@{-}[ur]_{#3} & & & \ar@{-}[ul]^{#4}\\}
}

\newcommand{\monogon}[1]  
{
\xymatrix{ *{\bullet}
\ar@
{-}
@(ul,dl)
[]
_{#1}
\\}
}

\newcommand{\bigon}[2]  
{
\xymatrix{ *{\bullet}
\ar@{-}
@/^1pc/
[r]
^{#1}
\ar@{-}
@/_1pc/
[r]
_{#2}
&
*{\bullet}
\\}
}

\newcommand{\thetagraph}[3]  
{
\xymatrix{ *{\bullet}
\ar@{-}
@/^1.5pc/
[r]
^{#1}
\ar@{-}
@/_1.5pc/
[r]
_{#2}
\ar@{-}
[r]^{#3}
&
*{\bullet}
\\}
}

\newcommand{\unigon}[1]  
{
\xymatrix{ *{\bullet}
\ar@{-}
[r]^{#1} 
&
*{\bullet}
\\}
}

\newcommand{\fourtheta}[4]  
{
\xymatrix{ *{\bullet}
\ar@{-}
@/^1.5pc/
[r]
^{#1}
\ar@{-}
@/_1.5pc/
[r]
_{#2}
\ar@{-}
@/^/
[r]^{#3}
\ar@{-}
@/_/
[r]^{#4}
&
*{\bullet}
\\}
}

\newcommand\tenj{10J--symbol}
\newcommand\sixj{6J--symbol}
\newcommand\fifteenj{15J--symbol}
\newcommand\threej{3J--symbol}

\begin{document}

\title {A Lorentzian Signature Model for \\ Quantum General Relativity}
\author{ John W. Barrett }
\address{School of Mathematical Sciences, University
Park, Nottingham NG7 2RD, UK} 
\author {Louis Crane}
\address {Mathematics Department, Kansas State University,
Manhattan KS 66502, USA}

\begin{abstract} We give a relativistic spin network model for quantum
gravity based on the Lorentz group and its $q$-deformation, the Quantum
Lorentz Algebra. We propose a combinatorial model for the path integral given 
by an integral over suitable representations of this algebra. This
generalises the state sum models for the case of the
four-dimensional rotation group previously studied in \cite{BC}, gr-qc/9709028.

As a technical tool, formulae for the evaluation of relativistic spin networks for the Lorentz group are developed, with some simple examples which show that the evaluation is finite in interesting cases. We conjecture that the `10J' symbol needed in our model has a finite value.
\end{abstract}

\maketitle 
 
\section{Introduction}
In \cite{BC}, we proposed a model for quantized discrete general relativity
with a Euclidean signature.
The model was constructed by combining the structure of a certain tensor
category and of a full subcategory of it with the combinatorics of a
triangulated 4-manifold. The category was the representations of $\Uq\so(4)$,
for $q$ a $4n$-th root of unity and the subcategory the representations which
are called balanced or simple.

The main technical tool was the introduction of spin networks for these
representations, which we called relativistic spin networks. The
adjective relativistic is used because the relativistic spin networks
are related to four-dimensional geometry whereas the original $\SU(2)$ spin
networks of Penrose \cite P are related to three-dimensional geometry. The name anticipated the development of relativistic spin networks for the physically realistic case of the Lorentz group, $\SO(3,1)$. The wider context for the model is explained in \cite{BZ, BZ3}. It is closely related to topologically invariant models \cite{CKY, W3D, TV, PR, BQG}.

The purpose of this paper is to supply the analogous concepts to our previous work for the Lorentz group and
its $q$-deformation, $\Uq\SL(2,\C)$. The geometrical description of the relativistic spin network evaluation for $\SO(3,1)$ is developed, in which hyperboloids in Minkowski space replace the role played by the sphere $S^3$ for the $\SO(4)$ case \cite {BC, B, FK}. 

We develop a correspondence between relativistic spin networks and the Lorentzian geometry of simplicies. One of the predictions we make is that the area of a timelike surface is quantized but the area of a spacelike surface can take any value in a continuous range.

After removing a trivial infinite factor of the volume of the hyperbolic space,
we obtain a expression for the evaluation of the relativistic spin network associated to a 4-simplex as a multiple integral on hyperbolic space. We conjecture that this integral is absolutely convergent. 

The study of the $q$-deformation of the Lorentzian case is not so far advanced, but we give some observations on this and the possible implications for 
Lorentzian signature state sum models. 

Before explaining our results, we give a brief review of the Euclidean case.

\subsection {Review of the Euclidean model}

The steps of the construction of the model in \cite{BC} were as follows:

\smallskip
\begin{enumerate}
\item Describe the geometry of a discrete Riemannian triangulated 4-manifold by
assigning bivectors to the 2-simplices satisfying appropriate constraints.

\item Identify the bivectors with Lie algebra elements.

\item Quantize the bivectors by replacing the Lie algebra with a sum over its
representation category.

\item Implement the constraints that the bivectors are simple bivectors by 
passage to a subcategory.

\item Switch to the representations of a quantum group with $q$ a root of unity
in order to create a finite model.

\item Determine a quantum state for each tetrahedron. This is a morphism
intertwining the four representations which are associated to the four boundary faces of the tetrahedron.

\item Connect the representations and morphisms around the boundary of each
4-simplex into a closed diagram called a relativistic spin network. The diagram consists of one tetravalent vertex at the centre of each tetrahedon with edges corresponding the four faces of the tetrahedron. Edges corresponding to the common face of two adjacent tetrahedra are then joined. This gives a closed graph embedded in the boundary of the 4-simplex, $S^3$. The evaluation of this relativistic spin network determines the amplitude for a single 4-simplex.

\item Multiply the amplitudes for each 4-simplex in the space-time manifold
together, and then sum over the representation labels introduced to give a
discrete version of a path integral.
\end{enumerate}

We introduced two variants of the model, given by different relativistic spin
networks for the 4-simplex. One of them is based on a \fifteenj\ from representation theory (a complex number which is a function of 15 irreducible representations). The other is based on a new function of 10 irreducible representations which one might call the \tenj\footnote{despite the fact that it is not one of the traditional 3NJ--symbols}.

Firstly we describe the model based on the \tenj.
There is a uniquely determined morphism which
satisfies all the constraints for the tetrahedron, which is taken to be a
4-valent vertex for the relativistic spin network. The vertex was introduced in
\cite{BC} and generalised to other valencies in \cite Y. For the $q=1$ case,
the vertex was proved to be the unique one satisfying the constraints for the
geometry of a tetrahedron in \cite R, and its detailed properties were
investigated in \cite{Barb2,BB}. This vertex can be regarded as the quantum state determined by quantizing the four bivectors determined by the four faces of a classical geometrical tetrahedron.

Starting with this 4-valent vertex one obtains an
amplitude for the 4-simplex which depends on the 10 representation labels at
the 10 triangles. The irreducible representations in question are indexed by the integers, and the geometrical interpretation of these is that they are the areas of these 10 triangles in suitable units. The asymptotic properties of this amplitude are related to the
Einstein-Hilbert Lagrangian in \cite {B},\cite{BWA}, and the classical
solutions for a simplicial manifold are determined in \cite{BRW}. We refer to the expression formed from the tetravalent diagram on a 4-simplex as a \tenj, and the model constructed from them as the 10J model.

For the 15J model one takes the relativistic spin network 
$$\DoubleY{}{}{}{}{b}$$
obtained by composing two trivalent vertices as the quantum state for the
tetrahedron. This depends on the additional choice of a simple/balanced representation $b$ for the centre edge.

The closed relativistic spin network for the 4-simplex is a trivalent graph
called a \fifteenj.  The asymptotic properties of the \fifteenj\ are
related to the Einstein-Hilbert Lagrangian in \cite{CYA}. The principal
difference to the 10J model is that the amplitude for the 4-simplex
depends on 15 irreducible representations, not 10. For each tetrahedron, one has
a representation on each triangular face, together with a fifth representation associated to the tetrahedron. To evaluate the \fifteenj\ one requires a choice of splitting the four faces into two pairs.

The geometrical picture of the 15J model in terms of bivectors is a little more complicated than the 10J case, as there are now five bivectors for a tetrahedron instead of four. Four of the bivectors are now associated to the faces of a combinatorial tetrahedron, but they no longer satisfy the geometrical constraints for the bivectors of a tetrahedron. The constraints that these satisfy are that the
bivectors for one pair of faces and the fifth bivector lie in a hyperplane
$H_1\subset\R^4$, and the bivectors for the other pair of faces and the fifth
bivector lie in a second hyperplane $H_2$. This is clear by carrying out the
analogous geometrical analysis to that in \cite{B}; each vertex of a closed
relativistic spin network is associated a unit vector $n\in\R^4$ and an edge is
associated the bivector $* (n_1\wedge n_2)$. The fact that there are two
hyperplanes for the tetrahedron and not one is not accidental; the variable $b$
is canonically conjugate to the angle between the two hyperplanes, so that if
$b$ is specified precisely then the angle is necessarily indeterminate, due to
the uncertainty principle.

The trade-off between these two versions of the model is that introducing the
extra representation on the tetrahedron increases the coupling between
neighbouring 4-simplexes in a state sum model but destabilises the geometry of
the 4-simplex. At present the best we can say is that the interpretation of the
model is unclear at this point, but the clarity is neither better nor worse in
its Lorentzian version.

\subsection{Results in Lorentzian signature}

The purpose of the present paper is to propose analogous models for the
Lorentzian signature. As we will explain below, this essentially amounts to
replacing $\Uq\so(4)$ with the noncompact quantum group $\Uq\sll(2,\C)_\R$, (known in
the literature as the Quantum Lorentz Algebra) whose representations have been
studied in \cite{PUSZ, BR}. Some new phenomena occur due to the fact that the
Lorentz group is not compact. For example, it seems no longer to be necessary to set the quantization parameter $q$ equal to a root of unity, instead, we conjecture that any real number will do.

Unfortunately, the representation categories of noncompact Lie groups or
quantum groups are more complex objects than the categories which have appeared
in constructions of TQFTs or Euclidean general relativity. In order to
formulate categorical state sums for these new categories, we will eventually
(although not here), find it necessary to
formulate the axioms for a new class of tensor categories to be called
``measured categories''. The new class of categories is a generalization of the
theory of
unitary representations of noncompact groups due to Mackey \cite{MA} and of the notion of fabrics due to Moussouris\cite{MOU}.

For the purposes of this paper, the difference is that the irreducible
representations are not discrete, but form a measure space.
This leads to a natural generalization of a state sum
which we call a state integral.

It is not quite possible to give explicit formulae for our model with the
present stage of knowledge of the representation theory of the Quantum Lorentz
Algebra. The progress we have made is as follows

\begin{enumerate}\item
There is a natural definition of the \tenj\ for $\SO(3,1)$ which we conjecture to be finite. 

\item The integration measure for the state integral is a finite measure for the $q$-deformed case, the QLA.
\end{enumerate}

In the case of 1.~we present some arguments for this conjecture.

Obviously one would like to conjecture that the \tenj s remain finite for the QLA. However we do not yet have the correct definition of the $q$-deformation.
Also, the question of whether there is a good definition of the \fifteenj\ for $\SO(3,1)$ and for the QLA is not yet clear.

The following is an outline of the rest of this paper: chapter 2 gives a review
of the representation theory of $\SO(3,1)$. Chapter 3 explains the geometry of
relativistic bivectors and its relationship to representation theory and
quantization. Chapter 4 deals with the question of the vertex for the theory.
Chapter 5 discusses the state sum model for the theory in the divergent case of
the classical Lorentz group, giving details of the definition of the Lorentzian relativistic spin network evaluation.
Chapter 6 recalls the basic facts about the representation theory of the
Quantum Lorentz Algebra and
proposes a $q$-deformed model with integrals with respect to a finite measure.
  Finally, chapter 7
proposes some natural directions for the investigation of the model.

\section{The representation theory of $\SO(3,1)$}

The purpose of this brief chapter is to acquaint the reader with the necessary
facts about the representation theory of the Lorentz group and its Lie algebra,
the Lorentz algebra \cite{6,7,8}.

The history of the subject is actually rather strange. It was
originally studied by Dirac, who thought that representations of the Lorentz
group, which he called ``extensors'' \cite{Dirac} might play a role in physics.
In the subsequent development of quantum field theory, the representations
of the Poincar\'e group which were important were not the ones which came from
the Lorentz group, and physical interest in the subject disappeared (at least
until the present). On the other hand, the work of Gelfand and Naimark
\cite{6,7,8}
was extremely influential in mathematics.

In this paper we are only interested in the unitary
irreducible representations of the Lorentz algebra called the
principal series. We use unitary representations in order to have a probability interpretation. Existing state sum models use the Plancherel measure in the summation over representations. In the Plancherel measure for the Lorentz group the irreducibles which are not in the principal series have measure zero and can therefore be neglected.

The principal series representations are labelled by two parameters, one a half integer $k$,
the other a continuous real parameter $p$; we can denote the representations
$R(k, p)$. The two parameters are naturally combined into a single complex
number, $w=k+ip$.
There are no isomorphisms among them except that $R(k, p)= R(-k,-p)$. Each is
irreducible, infinite dimensional and isomorphic to its dual. Any unitary
representation of the Lorentz algebra can be written as a direct integral of
irreducibles. The regular representation can be written as a direct integral
involving only the principal series.

Representations of the principal series are classified as fermions or bosons
accordingly as $k$ is half integral or integral. Both determine representations
of the covering group of the Lorentz group, which is the group $\SL(2,\C)$
considered as a real Lie group. The bosonic representations give
representations of the Lorentz group itself. The tensor product of any two
members of the principal series is a direct integral of all the bosonic members
if both of the representations are bosons or both fermions, and a direct
integral of all the fermions otherwise.

The Lorentz algebra can be thought of as $\sll(2,\C)$ considered as a real Lie
algebra, or as $\so(3,\C)$ considered as a real Lie algebra. This corresponds to
writing a general element of the Lorentz algebra as a sum of a rotation $J$ and
a boost $K$. Then $J+iK$ is the corresponding element of $\so(3,\C)$.
There are two invariant inner products on this Lie algebra,
$$\langle L,L\rangle ={\frac12} L_{ab}L^{ab} =J^2-K^2$$
and
$$\langle L,*L\rangle ={\frac14} L_{ab}L^{cd}\epsilon^{abcd}=2J\cdot
K$$
The corresponding Casimir elements in the Lie algebra have eigenvalues
$$C_1=k^2-p^2-1$$
$$C_2=2kp$$
These are the real and imaginary parts of $w^2-1$.

\section {Quantizing simple bivectors}

The space of bivectors over $\R^4$ is a six dimensional real vector space.
If we equip $\R^4$ with a Euclidean metric, then the bivectors have a natural
identification with the Lie algebra $\so(4)$, which was necessary in the
construction in \cite{BC}. As explained in \cite{B, BB}, the correct
isomorphism is to take $b=*L$, or
$$ b^{ab}={\frac12}\epsilon^{abcd}{L^e}_d g_{ec}$$ with $g_{ec}$ the Euclidean metric.
If take $g$ to be  a Lorentzian pseudometric on $\R^4$ (Minkowski space) instead of the Euclidean metric, then the bivectors are
naturally identified with $\so(3,1)$ instead of $\so(4)$. Although changing the signature of the Lorentzian psuedometric would introduce a minus sign into the above identification, all bilinear conditions on the bivectors are unaffected.

The bivector associated to a triangle is the wedge of two edges, a {\it simple}
bivector.
The condition that the bivector is simple is
$$\langle b,*b\rangle=0,$$
 the same equation as for the Euclidean case.

In addition, we would like to differentiate spacelike, null and timelike simple
bivectors, which correspond to planes in Minkowski space with an induced metric
which is Euclidean, degenerate or Minkowskian. These are determined by the sign
of $\langle b,b\rangle$.
This is positive for spacelike bivectors, zero for null bivectors and negative
for timelike bivectors. However the Hodge $*$ interchanges timelike and
spacelike bivectors, so the corresponding Lie algebra element $L$ has
$\langle L,L\rangle$
negative for spacelike bivectors, zero for null bivectors and positive for
timelike bivectors.

The idea behind the construction of this paper, as well as the model in
\cite{BC},
is that having transformed all the variables for discretized GR into the form
of constrained angular momenta (equals bivectors thought of as infinitesimal
rotations) we quantize them the same way we quantize ordinary angular momenta,
by  replacing them with representations of the appropriate Lie algebra. Perhaps
this is slightly obscured by the fact that in three dimensions bivectors are
Hodge dual to vectors, so nonrelativistic angular momentum is written as a
vector operator.

In this program, it is now desirable to find an expression of the constraint
that a bivector be simple translated into the category of unitary
representations of the Lorentz algebra.

The condition that the bivector is simple, $\langle b,*b\rangle=0$, translates
into the vanishing of the corresponding Casimir $C_2=2kp$. Thus either $k=0$ or
$p=0$. The quantization of $\langle L,L\rangle$ is then $C_1+1=k^2-p^2$.
If $p=0$, then $\langle L,L\rangle$ is positive, and so the Lie algebra element $L$ is spacelike. This means that the bivector $b=*L$ is timelike. This
leads us to propose the subcategory of representations $R(k,0)$ as the
quantization of the timelike bivectors, and the subcategory generated by the
$R(0,p)$
representations as a quantization of the spacelike ones.

Since the Hodge $*$ interchanges timelike and spacelike bivectors, the space of
simple timelike Lie algebra elements with a given square $\langle L,L\rangle$
is topologically the same as the corresponding space of spacelike elements.
However the Poisson structures are different. Indeed, the spacelike simple Lie algebra elements
include the subalgebra $\su(2)$, for which there is a
quantization condition that the symplectic form is integral. This leads to a
discrete series of representations. However the corresponding cohomology class
for the timelike elements vanishes, so there is no quantization condition in
this case and the parameter $p$ is arbitrary.

There is another physical motivation for labelling the faces with
the Hodge duals of their geometric bivectors. The approach we are exploring makes contact with the
spin foam proposal of Rovelli and Reisenberger \cite{RR}.

In that picture states for quantum gravity are  described by embedded spin
networks in 3d space. Their evolution is modelled by the world sheets of
evolving spin networks which can change in time by developing vertices. On
closer study, it seems that the kind of 2-complex they investigate is dual
to a triangulation labelled as a term in our state sum. Although their approach has self-dual $\su(2)$ connections, the natural extension to $\so(4)$ connections would contain bivector operators which are canonically conjugate to the connection variables. As explained in \cite{BB}, the Poisson structure which comes in here is the one determined by the Einstein action, and this identifies (dual) Lie algebra elements with bivectors using the Hodge $*$.

This contact with the
kinematics of what is superficially a very different approach to
quantizing gravity strengthens the specific proposal we have for the form
of our state sum, which was suggested by 
the facts of the representation theory of the Lorentz group.

\section{Vertices}

In this section we consider the possible forms for the vertices for
relativistic spin networks by considering the case $q=1$, in which 
the constructions can be described geometrically.

Naimark \cite7 has shown that between any triple of representations of
$\SL(2,\C)$
satisfying the parity condition there is a unique (up to scalar multiple)
intertwining operator. This means that, apart from normalisations, the
trivalent vertex for the relativistic spin networks is defined. 

To obtain a picture for the tetrahedron, we consider the possible
4-valent vertices. In the case of $\SO(4)$, the requirement that each tetrahedron lies in a
hyperplane (3-plane) translated into the constraint that the sum of any pair of
bivectors on two faces of a tetrahedron add up to form another simple bivector.
This is still true in the Lorentzian case.

However it is not true that if the bivectors on the faces are spacelike then the sums of pairs will necessarily be spacelike as well. Likewise if the faces are timelike then the sum of any pair of bivectors need not be timelike. These facts are mirrored in the representation theory: tensor products of two even-spin 
$R(k,0)$ representations will contain copies of the $R(0,p)$ representations and vice versa. This will need to be taken into account below, when we decide exactly how to define a state sum model.

Our preferred representations $R(k,0)$ and $R(0,p)$ can be realised in the
spaces of square integrable functions on the hyperboloids in Minkowski space,
$\R^4$ with inner product $x\cdot x=(x^0)^2-(x^1)^2-(x^2)^2-(x^3)^2$, using the Gelfand-Graev transform \cite{GG, GGV}. We consider the cases $Q_1$ given by $x\cdot x=1$,
$x^0>0$, the positive null cone $Q_0$ given by  $x\cdot x=0$, $x^0>0$ and the
de-Sitter space $Q_{-1}$, given by $x\cdot x=-1$. In \cite{BC}, we
noted that the simple/balanced representations of $\SO(4)$ could be realised in the
space of functions on $S^3$.  The relativistic spin network formalism in terms
of functions on $S^3$ was developed in \cite{B, FK}. This can be directly
generalised to the present case in a number of ways by replacing $S^3$ with either one of the hyperboloids $Q_1$, $Q_0$ or $Q_{-1}$. 

The representations of $\SO(3,1)$ have a Fourier decomposition
as the following direct integrals and direct sums:
$$ L^2(Q_1)\cong\bigoplus_p R(0,p) \dd \mu_p$$
$$ L^2(Q_0)\cong2\bigoplus_p R(0,p) \dd \mu_p$$
$$ L^2(Q_{-1})\cong\left(2\bigoplus_p R(0,p)\dd \mu_p\right)\bigoplus\left(\bigoplus_k
R(k,0)\right). $$
The notation $\oplus \dd \mu_p$ indicates a direct integral, using the measure $\dd \mu_p=p^2\dd p$, which is the Plancherel measure restricted to $k=0$. 
An element of the
direct integral is a vector function of $p$, $v_p\in R(0,p)$, with square
$\int\|v_p\|^2\dd \mu_p$. The 2 indicates that the following term appears as a
summand twice.

This decomposition can be understood in the following way. The Casimir operator
$C_1$ is the Laplacian on $Q_1$. So the irreducible representation $R(0,p)$ can
be considered the space of solutions to the eigenvalue equation for the
Laplacian with eigenvalue $-1-p^2$.

The Casimir $C_1$ is also the wave operator on $Q_{-1}$ with
eigenvalues either $-1-p^2$ or $k^2$. In this case one has solutions of the wave equation
associated to both the time orientations, somewhat analogous to positive
energy and negative energy solutions of the wave equation in Minkowski space.
This gives the two
copies of $R(0,p)$ in the Fourier decomposition of $L^2(Q_{-1})$\cite{V}. 
In addition there are the `tachyons', giving the $R(k,0)$. 
For the hyperboloid $Q_0$, one has a degenerate version of these formulae.

One would like to have an intuitive understanding of these Fourier decompositions in terms of the quantization of bivectors. Bivectors in Minkowski space determine geodesics on the hyperboloids. For example, a spacelike simple bivector is Hodge dual to an oriented timelike plane through the origin in Minkowski space, which intersects $Q_1$ in an oriented geodesic. Following Mukunda \cite{M}, quantum mechanics on $Q_1$ has as a classical limit the motion of free particles on $Q_1$. These move along geodesics given by the Hamiltonian $H=P^2$ on the phase space $T^*Q$, where $P$ is the
coordinate for the cotangent space. In fact, the constraint surface $\{H={\rm constant}\}$  decomposes
into the orbit space under this flow, namely the space of geodesics on $Q_1$, or the space of simple spacelike bivectors in Minkowski space.

  Each timelike plane intersects $Q_0$ and $Q_{-1}$ in {\it two} timelike or null geodesics each, one future-directed and one past-directed. This explains the multiplicity 2 of the $R(0,p)$ in the Fourier decompositions of these spaces.

Likewise, a timelike simple bivector is Hodge dual to a spacelike plane through the origin in Minkowski space, which intersects $Q_{-1}$
in exactly one spacelike geodesic. These planes do not intersect $Q_0$ or $Q_1$. Thus the $R(k,0)$ representations occur only in the Fourier decomposition of $Q_{-1}$. In this way, we understand the
multiplicities in the above Fourier decompositions.

The Gelfand-Graev transform gives precise formulae for the decompositions. The case of the 
three-dimensional hyperbolic space $Q_1$ is particularly important in the following, so we 
develop the formulae here. The other cases have analogous formulae.

In general the representations $R(k,p)$ can be realised in the space of square-integrable 
sections of a line bundle over $\C P^1$. However for the $R(0,p)$ one can simplify this to 
give the representation in the space of homogeneous functions of degree $-1+ip$ on the 
light cone in Minkowski space.\footnote{Gelfand and coauthors use the parameter $\rho=2p$.} 
This means an element $f$ of this space satisfies $f(\lambda \xi)=\lambda^{-1+ip}f(\xi)$ 
for null vectors $\xi$ and for any real number $\lambda\ne0$. 
The inner product on this space is determined by the integral
\begin{equation}\label{innerproduct}
\int_\Gamma \bar f_1(\xi) f_2(\xi) \dd\xi
\end{equation}
over the two-sphere $\Gamma$ given by the null vectors satisfying $\xi^0=1$.
The measure is the standard rotationally-invariant measure $\dd\xi$, normalised to total volume $1$.
The Gelfand-Graev transform gives the function
$$ \hat f(x)=\int_\Gamma f(\xi) (x\cdot\xi)^{-1-ip}\dd\xi$$
defined on $Q_1$.

Our proposal for a $k$-valent relativistic spin network vertex for $\SO(3,1)$ is given by
a map $R(0,p_1)\otimes\ldots\otimes R(0,p_k)\to\C$ by the formula
$$ f_1\otimes f_2\otimes \ldots f_k\mapsto {\frac 1{2\pi^2}}\int_Q \hat f_1(x) \hat f_2(x) \dots \hat f_k(x)\dd x.$$
In this formula, $Q$ is one of the hyperboloids $Q_{\pm1}$, $Q_0$, and $\hat f$ denotes one of the representations of $f$ as a function on the hyperboloid, as given for 
example for the case of $Q_1$ by the preceding formula. The measure $\dd x$ is the standard Riemannian (or pseudo-Riemannian) volume measure on $Q$. The map is not defined on every element of the tensor product, and the important point is to understand it as a generalised function of the $p_i$ variables as well as the $\xi_i$.

Alternatively the vertex can be defined by the formula for the kernel of the integral. For the case of $Q_1$ this is
$$V(\xi_1,\xi_2,\ldots,\xi_k)=\nfactor\int_{Q_1}
(x\cdot\xi_1)^{-1-ip_1}(x\cdot\xi_2)^{-1-ip_2}\ldots(x\cdot\xi_k)^{-1-ip_k} \dd x$$

Each of the representations for the different hyperboloids gives rise to a particular formula for the vertex
(of arbitrary valence) for the relativistic spin networks. Using $Q_1$, there is a single vertex for relativistic spin networks labelled
with $R(0,p)$. The uniqueness is due to the fact that each irreducible appears
only once in the
decomposition of $L^2(Q_1)$. Remarkably, this also implies a decomposition of
the four-valent vertex
\begin{equation}\label{yy}
\vcenter{\FourX{a}{b}{c}{d}}
=\int_p \dd \mu_p
\vcenter{\DoubleY{a}{b}{c}{d}{p} }
\end{equation}
with the intermediate edge labelled with the $R(0,p)$. This formula is
obtained by applying the Fourier decomposition to the intermediate product
function $\hat f_1(x)\hat f_2(x)$. As this is also a function on $Q_1$, its decomposition only involves the $R(0,p)$. Precise formulae and a proof of this decomposition are given below in section \ref{evaluation}. This formula is a direct analogue of the corresponding formula for the Euclidean case which formed the original definition of the four-valent relativistic spin network vertex in \cite{BC}.
The analogy is given in section \ref{comparison}.

Our interpretation of this formula is of a tetrahedron which lies in a
spacelike hypersurface. In this situation, the sum of the two bivectors for two
faces is always again a simple spacelike bivector. This is reflected in the
fact that the decomposition formula only requires intermediate representations
of the form $R(0,p)$.

Using $Q_{-1}$, there are a number of vertex formulae given by analogous
integrals. If the representation on a free end
is $R(0,p)$, then one has to specify whether the representation is to be
realised in $L^2(Q_{-1})$ as the future- or past-directed solutions of
the wave equation. These two transforms are given by
$$ \hat f(x)=\int_\Gamma f(\xi)G^\pm(x,\xi)\dd\xi$$
with $$G^+(x,\xi)= \left\{
\matrix(x\cdot\xi)^{-1-ip}& x\cdot\xi>0\\
0& x\cdot\xi<0\endmatrix \right.
$$
and $G^-(x,\xi)=G^+(-x,\xi)$\footnote{This differs from the integral transform given in \cite{GGV}, which is valid for functions which are even under inversion.}
Also, it is possible to put the $R(k,0)$ on the free ends, in only one way, using the formulae in \cite{GGV}.

Our interpretation of these 4-valent vertices is that they represent
tetrahedra which lie in a timelike (Minkowski signature) hypersurface in Minkowski space.
Accordingly the faces of such a tetrahedron can be timelike, either
future-pointing or past pointing, or spacelike. These possibilities correspond
to the different possible vertex formulae. There is also a
decomposition formula analogous to (\ref{yy}) for this case, which entails the use of two copies of $R(0,p)$ (the choice of $G^+$ or $G^-$) and the use of the
$R(k,0)$ representations. This corresponds to the fact that in a Minkowski signature tetrahedron the
sum of the bivectors for a pair of faces can be spacelike (future or past oriented) or timelike.

Finally, the analogous formulae for $Q_0$ give null tetrahedra.

 \section{The state sum model}

We shall now choose a geometric form for the model. Let us assume that the classical geometry is a triangulation of our manifold into 4-simplices all of whose boundary tetrahedra are spacelike. This means that all of the bivectors on the 2-simplices and all of the sums of bivectors of 2-simplices in the boundary of a common tetrahedron must be simple and spacelike. As discussed in the previous section, this is consistent with a categorical calculus in which the
representations $R(0,p)$ only are used everywhere. This is the simplest form
for a model, though we note that it may be interesting to investigate the other
possibilities.

For the 10J version of the model, a simplicial manifold is labelled with a
value of $p$ on each triangle, and the weight for this state is the product of
the symbols for each 4-simplex. In the 15J version of our model there is
additionally a value of $p$ on each tetrahedron, and we use the product of the
\fifteenj s. The model is formally the integral with the measure $\dd \mu_p$ on each variable in the simplicial manifold.

Of course, a model constructed from representations of $\sll(2,\C)$ would
involve either an integral over all values of $p$ (or for other forms of the model a sum over all values of $k$). This
would make it divergent. In \cite{BC} we solved the analogous problem by
passing to a quantum group. We shall see in section \ref{qla} how this seems to work out in
the Lorentzian case. In this section we state the form of the model for $q=1$.

\subsection{Relativistic spin network evaluation}\label{evaluation}

Firstly there is the question of how to evaluate the categorical
diagrams we are to associate with the 4-simplices. 

Given a function $h\in L^2(Q_1)$ the $p$-th irreducible component is given by
$$ f_p(\xi)=\nfactor\int_{Q_1}h(x)(x\cdot \xi)^{-1+ip}\dd x$$
giving a homogeneous function on the light cone, $\xi\cdot\xi=0$.
The inverse transform is
$$ h(x)=\int_0^\infty\dd\mu_p\int_\Gamma f_p(\xi) (x\cdot\xi)^{-1-ip}\dd\xi.$$
Composing these two transforms gives the delta function $\delta(x,y)$ on $Q_1$. However if one does the same calculation but does not integrate over $p$, the result is a projection operator onto the $p$-th irreducible component $h_p$ of the function $h$ on $Q_1$. The kernel of this operator can be explicitly evaluated. The projection is given by
$$    h_p(x)=\nfactor\int_{Q_1}K_p(x,y)h(y)\dd y$$
where 
\begin{equation}\label{zonal}
K_p(x,y)=\int_\Gamma (x\cdot\xi)^{-1-ip}(y\cdot\xi)^{-1+ip}\dd\xi
=\frac{\sin pr}{p\sinh r},
\end{equation}
$r$ being the hyperbolic distance (boost parameter) between $x$ and $y$. For a fixed $x$, the function of $y$ is called the zonal spherical function \cite{V}. It is the solution of the Helmholtz equation which is spherically symmetric about $x$.

One can check explicitly that the delta function is regained by integrating over $p$
\begin{equation}\label{delta}
\int_0^\infty K_p(x,y) \dd\mu_p=2\pi^2\delta(x,y)
\end{equation}

For a fixed $x$, $K_p(x,y)$ gives a continuous function of $y$ which is absolutely bounded by the value $1$ at $x$ and decays exponentially at infinity, but is not square integrable. The normalisation is in fact
$$ \nfactor\int_{Q_1} K_p(x,y) K_{p'}(y,z) \dd y = K_p(x,z)\frac{\delta(p-p')}{p^2},$$
expressing that fact that $K$ is a projection to the $p$-th Fourier component on $Q_1$. Here it is assumed that $p, p'>0$.

Now we can give the rules for evaluating a relativistic spin network. 
The network is a graph with each edge labelled by a parameter $p$. The graph is allowed to have a boundary. This means that the edges do not have to end in vertices of the graph. The ends of the edges which do not meet vertices are called free ends, and the set of all these is the boundary of the graph. If there are no free ends the graph is called a closed graph. 

The naive evaluation associates one variable $\xi$ to each edge and
is given by taking the product of one kernel $V(\xi_1,\ldots,\xi_k)$ for each vertex and integrating along the interior edges (those without free ends) using the inner product (\ref{innerproduct}) in the $\xi$ variables.\footnote{The hermitian inner product (\ref{innerproduct}) is written as a bilinear product by replacing $p$ with $-p$ in one of the two factors.}

For the interior edges this procedure gives a factor of $K_p(x_1,x_2)$ for each edge.  This is because performing the transform from functions on $Q_1$ to $R(0,p)$ and back to the space of functions on $Q_1$ at the next vertex is the same as inserting $K_p(x,y)$ for the edge.

This can be illustrated by giving the promised proof of (\ref{yy}). This uses first equation (\ref{delta}), then (\ref{zonal}).
\begin{multline*}
\vcenter{\FourX{p_1}{p_3}{p_2}{p_4}}
=V(\xi_1,\xi_2,\xi_3,\xi_4)\\
=\int_{Q_1} \frac {\dd x}{2\pi^2}
\int_{Q_1}\frac {\dd y}{2\pi^2}
\int_0^\infty  \dd\mu_p\; 
K_p(x,y)
(x\cdot\xi_1)^{-1-ip_1}(x\cdot\xi_2)^{-1-ip_2}
(y\cdot\xi_3)^{-1-ip_3} (y\cdot\xi_4)^{-1-ip_4} \\
=\int_0^\infty  \dd\mu_p \int_\Gamma \dd\xi\; V(\xi_1,\xi_2,\xi)V(\xi,\xi_3,\xi_4)\\
=\int_0^\infty  \dd\mu_p \vcenter{\DoubleY {p_1}{p_3}{p_2}{p_4}p}
\end{multline*}

\subsection{The regularised evaluation for closed networks}

For closed networks the naive evaluation formula can be re-expressed entirely in terms of the kernels $K$. In this section we give the definition of the evaluation for closed networks using these kernels as we shall use it in the rest of this paper. It will be necessary to modify the naive evaluation formula a little to obtain the actual definition of the evaluation which we are to use.

Each vertex is associated a variable $x\in Q_1$.
Each edge is associated the factor $K_p(x_1,x_2)$, the variables being the ones associated to the vertices at either end of the edge. The naive evaluation is then
\begin{equation}\label{naive}
\int_{{Q_1}^n} \prod K_{p(ij)}(x_i,x_j) \dd x_1 \dd x_2 \ldots \dd x_n.
\end{equation}

 The evaluation of closed networks is a problem. Since the irreducible
representations are infinite dimensional, we cannot trace on them. This is because the trace of the identity operator is infinite.

In our integral definition this is reflected in the fact that the integral is invariant under $SO(3,1)$ and the orbits are not compact, so one has infinite factors for the volume of each orbit. 

There are two essentially equivalent ways of dealing with this.  The definition we adopt is simply to remove the integration over the variable in $Q_1$ at one of the vertices in \eqref{naive}. Using the Lorentz invariance, it does not matter which variable is chosen for this. We take the variable $x_1$ to be fixed.

\begin{defn}
The regularised evaluation for closed networks is 
\begin{equation}\label{closed}
\int_{{Q_1}^{n-1}} \prod K_{p(ij)}(x_i,x_j)  \dd x_2 \ldots \dd x_n.
\end{equation}
\end{defn}

Due to the Lorentz invariance, this integral is independent of the chosen value for $x_1$. This definition is similar to the suggestion
in \cite D that we can open up any closed network by cutting one edge to give a network with two free ends and regard the
network as an intertwiner from one irreducible to another. This gives a multiple of the identity operator. This multiple is then the evaluation of the original closed network.

The evaluation is illustrated by several very simple cases. 
Firstly, a graph given by a single loop with one vertex on it 
$$\monogon p$$
has the evaluation
$$ K_p(x,x)=1.$$

 A graph with two vertices on a loop 
$$\bigon ab$$
has the evaluation
$$ \nfactor \int_{Q_1} K_a(x,y) K_{b}(y,x)\dd y =\frac{\delta(a-b)}{a^2}.$$
This elementary example illustrates very clearly that the relativistic spin network evaluation may be distributional in the spin parameters $a,b$. This means that for particular values of these parameters (here $a=b$) the evaluation may not have a finite value.

Also we note that in the simpler case of a closed network with two vertices joined with just one edge, 
$$\unigon a$$
the evaluation is a divergent integral for all values of the spin parameter.
$$\nfactor \int_{Q_1} K_a(x,y)\dd y =\infty.$$

The theta symbol with two vertices and three edges 
$$\thetagraph abc$$
has the evaluation
\begin{multline*}\nfactor \int_{Q_1} K_a(x,y) K_{b}(x,y) K_c(x,y)\dd y\\
=\frac2{\pi abc}\int_0^\infty \frac{\sin ar\sin br\sin cr}{\sinh r}\dd r\\
=\frac1{4 abc}\bigl(f(b+c-a)+f(c+a-b)+f(a+b-c)-f(a+b+c)\bigr)\end{multline*}
where
$$ f(k)=\frac2\pi\int_0^\infty\frac{\sin kr}{\sinh r}\dd r=\tanh(\frac\pi2 k).$$

For large values of $|k|$, $f(k)$ tends to $\pm1$. As a consequence, when the Euclidean triangle inequalities for $a$,$b$,$c$ are satisfied the value of the theta symbol approximates $1/(2abc)$. When the inequalities are violated exactly one of the four $f$ terms becomes negative, giving approximately zero for the evaluation of the theta symbol in this limit. In fact the value dies away exponentially fast as $a$,$b$,$c$ increase away from the critical values of a degenerate triangle, characteristic of quantum effects in a classically forbidden region of configurations. 

In terms of bivectors this agrees with the interpretation developed above. The simple spacelike bivectors correspond to planes in Minkowski space which lie in a common spacelike hypersurface in Minkowski space. Whenever these three bivectors add to zero then their magnitudes satisfy the Euclidean triangle inequalities.

A similar analysis can be carried out for the graph with two vertices and four connecting edges. 
$$\fourtheta abcd$$
This graph is important because it gives a magnitude for the vertex for the tetrahedron. For this graph the evaluation is
\begin{multline*}\frac1{4 abcd}\Bigl(g(b+c+d-a)+g(c+d+a-b)+g(d+a+b-c)+g(a+b+c-d)\\
-g(b+c-a-d)-g(c+a-b-d)-g(a+b-c-d)-g(a+b+c+d)
\Bigr)\end{multline*}
where
$$ g(k)=\frac k2 \coth( \frac \pi2 k).$$
For large $k$, the function $g$ is asymptotically $|k|/2$. If one of four spins is bigger than the sum of the other three, say $d>a+b+c$, then the amplitude is asymptotically zero, just as for the theta symbol. However this condition occurs precisely when it is impossible to form a spacelike tetrahedron in Minkowski space with these numbers as the areas of the faces. This confirms our interpretation of the four-valent vertex as the quantization of a spacelike tetrahedron.

For more complicated graphs it is a non-trivial task to determine whether our expression for the evaluation is finite.
In the special case of a \tenj, we give arguments that the integral is 
finite and determines a function of the $p$. We do not at this point know a finite expression for a \fifteenj, or even a \sixj. If this is not just an artifact of our knowledge, it may be that only the 10J model survives in Lorentzian signature. The evaluation of relativistic spin networks is full of interesting open questions.

\subsection{Finite \tenj s  For $\SL(2,\C)$}\label{finite}
 
The \tenj\ is the evaluation of the relativistic spin network based on the complete graph on 5 vertices,
all of whose edges are labelled with $R(0,p)$ representations of $\SL(2,\C)$. 

\begin{conj} The regularised evaluation of the \tenj\ labelled with $R(0,p)$ representations is finite.
\end{conj}

{\bf Evidence for the conjecture:} This integral is similar in form to a Feynman integral on hyperbolic space. The kernel $K(x,y)$ is bounded as $x\to y$, so only infrared divergences need be considered. In other words, the only divergence possible would be at infinity in hyperbolic space. We therefore need to use two facts: the radial growth of the 
area of a large sphere in $Q_1$, and the asymptotic behavior at infinity of the kernel $K$. 

Using the fact that the area of the sphere is asymptotically $e^{2r}$, while the $K$ is asymptotically $e^{-r}$ it is possible to make an estimate for the integral expression (\ref{closed}). A delicate analysis shows that the expression is in fact absolutely convergent, even if the sine terms in $K$ are omitted and $K$ is approximated by $e^{-r}$ for large $r$.

Thus the behaviour of the integral (\ref{closed}) as all variables go to infinity separately is dominated by the integral of a positive function, and apart from the trivial factors of $1/p$, is independent of the values of the spin labels $p$ defining the representations on the edges.

In order to show that the integral is actually finite we need to consider also the regions where some vertices go to infinity and some do not. This corresponds to subdivergences in the standard Feynmanological language. We believe they do not appear in our case, but further analysis is needed.
  
It is interesting to note that the analogous analysis for a \sixj\ gives an indeterminate expression and it is not clear if the corresponding integral converges. It would be interesting to look for a regularisation procedure to give finite values for all diagrams in this category.

\subsection{Comparision with the Euclidean case}\label{comparison}

In our original paper on relativistic spin networks, we used the representation theory of $\SO(4)$ rather than $\SO(3,1)$. We call this the Euclidean case. In this, we used  irreducible unitary representations, which are finite dimensional. For these representations the spin network evaluations always exist as there is no difficulty taking the trace of a finite dimensional matrix. At first sight, the formalism for the infinite dimensional representations of the Lorentz group looks completely different. However the formulae we obtain have a strong similarity, and can say that the two evaluations are analogous but not the same. The irreducible representations that are used in the Euclidean case, the simple/balanced ones, are labelled by the spin $n$, an integer\footnote{This parameter is often given as a half-integer $j=n/2$.}. The following considerations show that the Lorentzian relativistic spin network evaluation can be considered as an extension of the Euclidean evaluation to the case where $n$ is a complex number of the form $-1+ip$, for $p\in\R$.

In \cite{B} the relativistic spin network evaluation for $\SO(4)$ was written in terms of integrals over $S^3$. This space is analogous to the hyperboloid $Q_1$ in the present work. Each edge is labelled with an integer $n$, and the kernel for each edge in \cite{B} is
$$K_E=\frac {\sin(n+1)\theta}{\sin\theta},$$ 
where $\theta$ is the distance between two points on $S^3$. 
On substituting $r=i\theta$ and $p=-i(n+1)$, the kernel $K$ for the Lorentz group becomes exactly $1/(n+1)$ times the formula for the Euclidean kernel $K_E$. This suggests that a better comparison between the magnitudes of Lorentzian and the Euclidean evaluations is to divide the later by a factor of $n+1$ for each edge.  Indeed this is confirmed by the example of the loop with one vertex. In \cite{B} the evaluation of the loop is $(-1)^n(n+1)$, whereas here it is $1$. 
With the adjustment, the evaluations agree up to a phase factor. In general understanding the comparison of the phase factors would require a deeper investigation.
 
For the theta symbol, the Euclidean evaluation gives $1$ if the three integer spin labels satisfy the triangle inequalities for a Euclidean triangle and sum to an even integer, and zero otherwise. For the Lorentzian evaluation, as computed above, the spin labels $a,b,c$ are continuous parameters and, after making the adjustment of multiplying by $abc$ (equivalent in magnitude to dividing the Euclidean formula by factors of $n+1$), the evaluation interpolates smoothly between asymptotic values of $1/2$ far into the interior of the region where the triangle inequalities are satisfied and $0$ far into the region where they are violated.

The integration measure is analogous too, as the factor of $1/(2\pi^2)$ which has been included with each integration on the unit hyperboloid is the volume of the unit three-sphere; so the analogous measure on $S^3$ is normalised to total volume $1$. This is the measure used in \cite B.

The decomposition formula (\ref{yy}) is directly analogous once one makes the correction of a factor of $n+1$ for the middle edge in the Euclidean evaluation. After this correction, the Euclidean formula would be to sum over the spins on the middle edge with weight $(n+1)^2$ (again ignoring phase factors). This is directly analogous to the measure $\dd \mu_p=p^2\dd p$ on substituting $n+1=ip$.

\section{Passing to the Quantum Lorentz Algebra: a finite model?}\label{qla}

The process of passing to the representation category of a quantum group should
not be thought of merely as a clever regularization scheme for this family of
models. Quantum groups fit very firmly into the program of noncommutative
geometry. A quantum group is a noncommutative space with a symmetry structure
like a Lie group. In fact, the noncompact quantum group which is referred to in
the literature as the Quantum Lorentz Algebra (QLA) has a good $C^*$ algebra
version \cite{BR,PW}. The set of irreducible representations of a $C^*$ algebra
is the noncommutative analog of the set of points of a space. Thus, the
construction of the model in this paper can be viewed as an exploration of a
noncommutative version of general relativity. The approach of this paper allows
us to
interpret the discoveries about the representations of the QLA as a species of
quantum geometry. In terms of classical physics, the $q$ deformation can also
be interpreted as the introduction of a cosmological constant \cite{Kod,BZ2}.

The representation theory of the QLA \cite{BR} is not as well understood as
that
of the classical Lorentz algebra. There are two possible forms of the QLA, one
with the deformation parameter real and one where it is a complex phase. It
seems that only the real case has been studied in the literature, so we shall
attempt to use it in our construction. As in the classical case, the
irreducible
unitary representations are infinite dimensional and classified by two
parameters, one discrete and one continuous. The first difference is that the
continuous parameter
is only allowed to take values in a bounded set of finite measure, which
depends on the discrete parameter.

The Plancherel measure (rather subtly defined) is given in \cite{BR} as
$$\frac {h}{2\pi} (\cosh(2hk)-\cos(2hp))\dd p,$$
where $q=e^h$. This measure is defined over $p\in [-\pi/h, \pi /h]$ and $k\in
Z$. However, since the values $(k,p)$ and $(-k,-p)$ are equivalent, the measure
is integrated over one half of this region, a fundamental domain for this
identification.

For $k=0$, the measure reduces to
$$\frac {h}{2 \pi}\left(1-\cos(2hp)\right)\dd p$$
over the interval $[0, \pi /h]$, which is a finite measure.

Now we can see that the effect of passing to the QLA is to make the
Plancherel measure restricted to the quantum version of the spacelike simple
bivectors a finite measure.

 On the other hand, the sum corresponding to the
timelike simple bivectors is not truncated. This is consistent with our choice
of model integrating over the representations of the form $R(0,p)$, thinking of
them as the Hodge duals of spacelike bivectors.
(At present we are actually constrained to do this because the version of the
QLA with a complex phase for $q$ is not studied. One could easily conjecture that
passing to a root of unity would give a truncation in $k$. This is a subject for
further study).

The rest of the situation with respect to the representations of the QLA
has not yet been fully clarified. The \threej s (trivalent vertices) have been defined only for a finite dimensional representation paired to a unitary one. On the other hand,
the universal R matrix is known, and Buffenoir and Roche have announced a
program to find the missing \threej s \cite{BR}.

\section{Prospects}

Clearly, the first item on the agenda in pursuing this model will be a careful
study of the 3J, 6J 10J and \fifteenj s of the QLA. Once the definitions of these symbols are clear, a number of natural questions will arise.

In the first place it will be interesting to see if asymptotic formulae for the relevant $q$-deformed
symbols can be found as in the compact case. This will allow us to see if the
argument recovering the Einstein Hilbert Lagrangian in the classical limit for
the Euclidean signature case can be extended to Lorentzian signature, as it
does in 2+1 dimensions \cite D.

Secondly, it will be necessary to study the state integral carefully to find out if the integrand contains any singularities which are not integrable.

Beyond this, there are many interesting questions we would like to study
about the model. One would like to see whether small disturbances in initial
conditions propagate causally.  The model differs from other quantum gravity models in that it allows a continuous spectrum for the area of a spacelike surface. It might be interesting to see if this can generate any interesting predictions for black hole spectra. 

Farther down the line, we would like very much to know what the model does
as we refine the triangulation on which it is based. Good behavior might tell
us how to use the model to construct an actual theory of quantum general
relativity.

\end{document}